# Quantum Monogamy with Predetermined Events

Ghenadie N. Mardari [1,2]

[1]  Open Worlds Research; Sparks, MD 21152, USA; gmardari@gmail.com

[2]  Rutgers University; Piscataway, NJ 08854, USA.

**Abstract**

The concept of correlation appears straightforward: measurement outcomes coincide, and patterns emerge. For any record of events, the coefficients are uniquely determined. Thus, if correlations change spontaneously, as seen in quantum monogamy, then individual behavior must have changed first. Surprisingly, this is not always true. When two observables are mutually exclusive, they cannot coincide objectively and need to be grouped across time. Yet, sectioning the flow of events into "iterations" is not trivial in this case. Even with blind windows of coincidence, the same order of outcomes can produce different coefficients of correlation, depending on the number of joint measurements. Therefore, quantum monogamy can happen with fixed pre-determined events. A new concept ("subjective correlation") is required to explain this phenomenon.

## 1. Introduction

Quantum monogamy is one of the most daunting puzzles in modern physics. It was discovered in the year 2000, when Coffman, Kundu and Wootters described quantum entanglement as a limited resource [1]. If a system A is maximally entangled with another system, system B, then it cannot also be entangled with a third system C. Shortly after, the concept was qualified and expanded by a suite of notable contributions [2–5], leading to a virtual explosion of research in this field [6,7]. Today, quantum monogamy is at the forefront of theoretical and experimental studies in quantum information theory [8], with special emphasis on protocols for quantum key-distribution [9–11]. It also had an impact on wider investigations into the basic nature of entanglement [12,13], condensed-matter physics [14], and even in black-hole physics [15]. Still, a fundamental obstacle to progress on this topic is the apparent inability to ex-plain its physical essence. An important aspect of monogamy is that a pattern of correlation between two observables can change spontaneously, just because another remote observable is added for joint measurement. This is a population-level effect that is not anticipated by event-level non-classical interpretations (e.g., non-locality, super-determinism, or retro-causality). Do we need to go even further into the realm of ontological speculation (perhaps with "everything, everywhere, all-at-once" determinism [16]), or should we go back to the drawing board of classical probability theory?

As currently understood, quantum entanglement is an irreducibly non-classical phenomenon. The search for intuitive explanations, in this context, is technically a search for loopholes [17]. This perception informed our previous contribution to this topic too [18]. For instance, quantum erasure requires post-selection, so a protocol with quantum monogamy should also apply to different sub-ensembles. Perhaps, triple co-incidences are less likely than double coincidences, and this is how different patterns are reconciled in a single system? Yet, a surprising recent discovery flipped the script on this project. As will be shown below, sequential physical properties cannot coincide with each other. The only way to achieve coincidences (and to derive correlations) is by combining them into pseudo-joint measurements *after detection*. Yet, the same record of events produces two different patterns of correlation, without changing the data in any way. The only factor that converts an "entangled" pattern into a "non-entangled" pattern is the decision to switch between pairwise and quadruple sectioning of events. Accordingly, the nature of physical reality prior to measurement cannot explain the features of quantum monogamy (and, implicitly, entanglement in general).

This is a shocking conclusion. We have at least one example where the appearance and disappearance of Bell violations happen after the act of measurement. By implication, there is no point in questioning Bell's Theorem [19], or the hypothetical loopholes of corresponding experiments [20–23], in order to avoid interpretive puzzles. The basic features of quantum entanglement are sufficiently explained by the incompatible nature of selected observables. Remarkably, this idea was previously explored by numerous scientists [24–36], but they could not identify a plausible physical mechanism for it. In practice, experimental data is expected to produce unique coefficients of correlation. If the pattern of coincidence changes, then it seems that the record of events must have changed too. As it turned out, mutually exclusive properties are immune to this rule of thumb. Even maximal ("super-quantum") Bell violations [37,38] are possible in classical systems, while also displaying the features of monogamy. To be clear, what is presented below is not a direct explanation of quantum behavior. Sequential properties are a special sub-class of incompatible variables with obvious differences from quantum spin (to give just one example). Yet, the textbook interpretation of quantum entanglement is based on a general argument about local behavior (Bell's Theorem), and it is this concept that is now cast in new light.

In what follows below, monogamous entanglement is described as a straightforward combinatorial problem with the help of a classical "wheel of fortune" toy model (Sections 2 and 3). A fixed record of events produces two types of coefficients automatically, depending on the number of observables that are included in the same window of coincidence. Maximal Bell violations are achieved with overlapping pairwise measurements. At the same time, no violations are possible with quadruple measurements. Conceptual implications are explained in Section 4 (as well as the Appendix A at the end), followed by more general comments and clarifications in Section 5. A notable conclusion is that quantum cryptography depends on the relational essence of incompatible coefficients. Its security does not require mystical resources and can even be improved with new tools that are likely to emerge from this discovery.

## 2.". Wheel-of-Fortune" Thought Experiment

A big obstacle in the analysis of quantum experiments is that we don't know what is going on. Do we measure properties that exist all at the same time, or one at a time? Do we have predetermined fixed events, or spontaneous manifestations? The advantage of a thought experiment is that such uncertainties can be removed by design. So, let us consider a system that is classical in every way, except that all the properties exist one at a time. To be clear, this is not like a system with four lights, where one light is always "on", and the other ones are "off". The correct analogy is a system that "shape-shifts" continuously, morphing sequentially into four different types of light emitters, one at a time. This is represented below with a "wheel-of-fortune" table that has a fixed order of manifestation for four properties (Figure 1). (This toy model is frequently used in probability theory and is based on a famous American TV show. Contestants spin the table and prizes become "real" when the corresponding sector stops under the arrow). In this case, the table displays a closed cycle of 8 sectors, with four positive and four negative values. As the table spins under the arrow, only one sector is "actualized" at a time. At this point, the value can be recorded with a time stamp. Therefore, joint observations for two properties can only be achieved across time, with extended windows of coincidence. Note that quantum experiments employ extended windows of coincidence to facilitate the identification of correlated pairs, and to minimize the probability of spurious counts. In contrast, this is an ideal classical system without uncertainty about the time of detection. The purpose of the window of coincidence is simply to force joint measurements for mutually exclusive events across time. Furthermore, quantum correlations are usually studied with remote measurements, to exclude the possibility of communication between stations. Yet here we have printed values that are fixed and cannot influence each other. It makes no difference if "Alice" and "Bob" events are on the same table, or on two identical tables. The advantage of a single table, as shown below, is that the relationship between events and correlations can be explored with visual clarity.

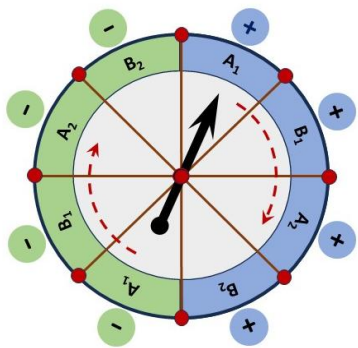

**Figure 1. "Wheel of fortune" toy model for visualizing fluid correlations between mutually exclusive properties.** The measurement outcomes of 4 binary variables (A1, B1, A2, B2) are arranged on the table in a closed cycle. The right half contains only positive values (highlighted in blue), with all the negative values on the left half (green). Each sector is actualized in order, as it passes under the tip of the black arrow. This corresponds to a detection event. As the events accumulate, one at a time, the same outcomes produce two different patterns of correlation, depending on the number of observables that are forced to coincide in a single window of coincidence.

The CHSH protocol, named after Clauser, Horne, Shimony and Holt [39], is the most common arrangement for a Bell experiment. It requires two observers, Alice and Bob, with the freedom to choose between two observables ($A_1$ and $A_2$ for Alice; $B_1$ and $B_2$ for Bob). The goal is to make a closed chain of overlapping joint measurements: $(A_1,B_1)$–$(B_1,A_2)$–$(A_2,B_2)$–$(B_2,A_1)$. As shown in Figure 1, this can be achieved by assigning dedicated axes with opposite values to each variable. "Alice" and "Bob" values are interlaced, such that any "Alice" event is placed between two different "Bob" events. An open question is whether a Bell violation is possible with pairwise consistency, such that the marginals of $(A_1,B_1)$ and $(B_1,A_2)$ always agree about the values of $B_1$, and so on around the chain. This locality criterion is achieved by default in this case, given the fixed sequence of predetermined values on the table. At the same time, event pairings cannot be random. The goal is to achieve the strongest possible correlations. This requirement is satisfied by placing all the positive-valued sectors on one half of the table, with the negative values on the other half. The table can be forced to rotate with constant speed, such that a fixed window of coincidence can always isolate joint observations from the recorded flow of events.

The most important feature of this system is that only one sector can be "real" (*i.e.*, pass under the arrow), at any point in time. This makes it impossible to have objective coincidences. This is why joint measurements require extended windows of observation. The width of the window of coincidence can be adjusted as necessary, to include two consecutive events or more. Furthermore, quantum measurements require preparation procedures. Here, the spinning table works as a mechanism with automatic sequential "preparations", such that any property can only manifest "when prepared", *i.e.*, when it gets a turn. The same cycle of events is repeated continuously, as long as needed for statistical significance. This makes it easy to infer the patterns of possible coincidence, since the sequence of measurement outcomes cannot change over time.

## 3.". Quantum Monogamy" with Fixed Measurement Outcomes

In a system with sequential properties, all the variables can have an equal chance of manifestation, in order. In this case, for every complete turn of the "wheel-of-fortune", every variable is expressed twice. Positive values for $A_1$-$B_1$-$A_2$-$B_2$ are followed by negative values in the same order. This means that a joint measurement of the four variables requires a window of coincidence that is equal to half the period of rotation. In principle, this procedure can be repeated indefinitely for data acquisition. Yet, the cyclic nature of this process entails a stable pattern of coincidence, with one possible actualization as shown in Figure 2(a). A complete record of events will have only two types of coincidences. Half the iterations will have positive events exclusively, as represented by the inner circle in the diagram. The other half will have negative values, as represented by the outer circle.

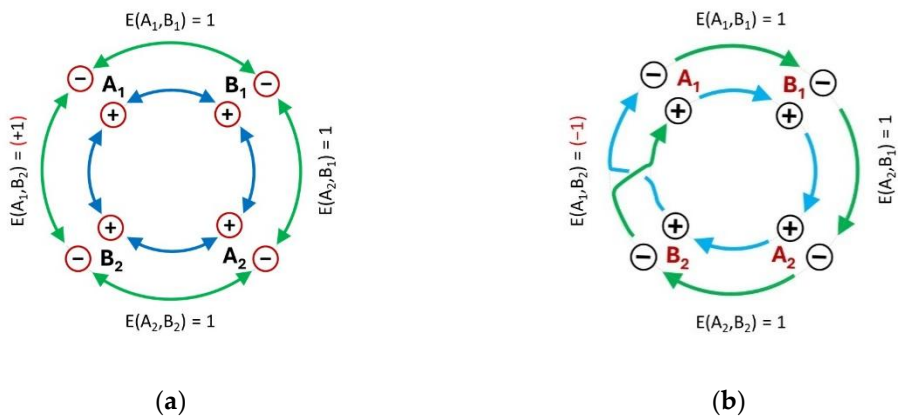

(**a**)          (**b**)

**Figure 2. Correlation patterns for different coincidence windows.** Arrow colors maintain the code from Figure 1, such that blue arrows originate on the positive-valued sectors, and green arrows on the negative. (a) Quadruple detection results in global joint distributions with four correlated pairs. (b) In contrast, pairwise detection produces an odd combination of three correlations and one anti-correlation. The underlying record of events is identical for both patterns.

Thus, no further analysis is needed to derive the four pairwise correlations for a CHSH test. The only possible combinations of events for the pair ($A_1,B_1$) are "+,+" and "−,−". Ergo, the expectation value is $E(A_1,B_1) = 1$. The same pattern holds for all the remaining pairwise combinations in the Bell chain. Accordingly, we can plug these coefficients into the CHSH inequality [39]:

$$|E(A_1B_1) - E(A_1B_2) + E(A_2B_1) + E(A_2B_2)| \leq 2 \qquad (1)$$

The output is a confirmation of the expected "classical" limit:

$$S = |1 - 1 + 1 + 1| = 2 \qquad (2)$$

Yet, there is more to this result than meets the eye. Classical objects tend to exhibit *permanent* combinations of simultaneous properties. It does not matter how many such qualities are measured at a time, because their relationships are objectively fixed, even in the absence of measurement. In contrast, here we have a system with mutually exclusive observables. The requirement of joint distribution [40] is satisfied conditionally. In other words, we have the appearance of joint distributions because we created it by force when four observables were chosen for simultaneous observation.

What happens if the window of coincidence is narrow enough to include only two events at a time? In this case, a full turn of the table corresponds to a double loop of a Mobius strip, as shown in Figure 2(b). Instead of batches of four events, where pairwise coincidences are extracted from within quadruple observations, we have a continuous suite of overlapping pairwise measurements. Surprisingly, this results in a "non-classical" combination of three correlations and one anti-correlation. Coincidences for ($A_1,B_1$), ($B_1,A_2$) and ($A_2,B_2$) remain the same as in the previous example, with nothing but "+,+" and "−,−" pairings. Therefore, they still have an expectation value of (+1). In contrast, coincidences between $A_1$ and $B_2$ happen across the two halves of the table (Figure 1), corresponding to the area with crossovers between the loops of the Mobius strip in Figure 2(b). The only possible combinations in this case are "+,−" and "−,+", with an expectation value of (−1). If these values are plugged into the CHSH inequality, we get a maximal Bell violation:

$$S = |1 - (-1) + 1 + 1| = 4 \qquad (3)$$

From a theoretical point of view, this is a legitimate outcome. Cyclic systems are known to exhibit consistent pairings without joint distributions, even in classical probability theory [40,41]. This makes them immune to Bell's Theorem [42]. Nonetheless, it is deeply surprising to see the patterns from Equations (2) and (3) in a single classical system, let alone *after* detection. The same flow of events produces two different combinations of expectation values, just as observed in the case of quantum monogamy. The only difference is that Bell violations are maximal in this case, leading to a general conclusion. Any kind of "monogamy" phenomena can be explained (at least in principle) with predetermined events.

As shown in Figure 3 below, this is a straightforward combinatorial effect. When all the properties are measured in a single window of coincidence, the flow of events is broken into non-overlapping batches of four events (Figure 3a). Every event is used twice, as required for a local Bell experiment. However, the values of $A_1$ are not paired with the closest manifestation of $B_2$. Instead, they are paired with a remote value of $B_2$ from the same coincidence window. In contrast, the narrower window of coincidence results in a continuous chain of overlapping pairwise measurements (Figure 3b). This means that $A_1$ and $B_2$ values are always grouped with the values that are closest in time to each other. Accordingly, switching from a quadruple joint measurement scheme to a double joint measurement scheme entails an unavoidable flip for just one out of four expectation values. This is enough to change the outcome of a CHSH calculation from a non-violation to a maximal violation.

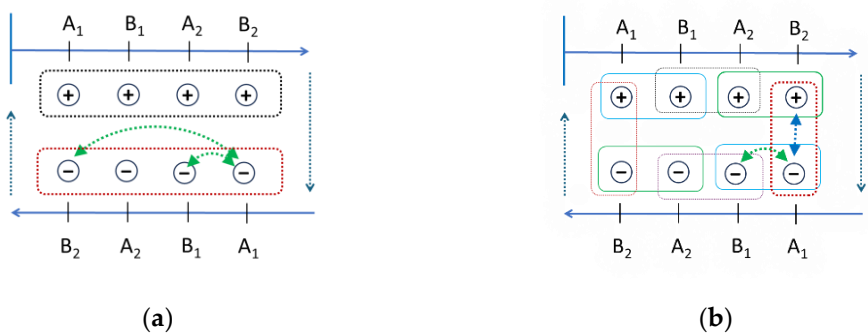

(a)          (b)

**Figure 3. Visual explanation of the "monogamy" mechanism.** (a) Quadruple detection forces pairwise combinations within each batch of four events. Two observables ($A_1$ and $B_2$) must be paired with non-adjacent events. (b) Pairwise detection allows all the events (including $A_1$ and $B_2$) to be paired with the nearest neighbor.

To sum up, mutually exclusive properties do not coincide objectively and therefore require artificial grouping. This process is sensitive to the number of variables that are combined into joint measurements, such that pairwise observations allow for Bell violations, while global observations do not. A fixed record of detection events is compatible with both patterns of coincidence that are associated with the so-called "quantum monogamy" phenomenon. This is only possible for mutually exclusive properties. Though, quantum theory is only known to predict this behavior for non-commuting variables. Therefore, this demonstration is sufficient to falsify the need for "new physics" as an interpretive resource for unusual quantum correlations.

## 4. Monogamy of Entanglement Explained

Quantum entanglement is measured by the ability of two systems to violate Bell's inequality [19]. To understand why this feature is "monogamous" (*i.e.,* restricted to pairwise coincidences), is to understand the root cause of this phenomenon. Hence, a quick review of the basic concepts is in order. What is the essence of Bell's inequality? The fundamental question is whether two observables can be physically independent, given that they are strongly correlated, and therefore statistically inseparable:

$$P(A,B) \neq P(A)P(B). \tag{4}$$

The classical answer to this question is Reichenbach separability [43,44]. If two observables have a common cause C, then they can be conditionally separable by virtue of such an additional parameter:

$$P(A,B|C) = P(A|C) P(B|C). \tag{5}$$

Bell's locality criterion is an extension of this principle to correlated systems that are space-like separated from each other, and therefore causally independent. If two observables from remote locations are *not* influencing each other directly, then their joint probability must by conditionally separable by virtue of a "hidden variable" $\lambda$:

$$P(A,B|\lambda) = P(A|\lambda) P(B|\lambda). \tag{6}$$

Notably, $\lambda$ is a catch-all term that includes any possible physical process that applies to both observables. The only restriction is that both parties should be determined by *the same* parameter $\lambda$. After all, observables A and B can also be separated with different hidden variables:

$$P(A,B|\lambda_A, \lambda_B) = P(A|\lambda_A) P(B|\lambda_B). \tag{7}$$

The problem is that now we need another prior cause, this time for the two hidden variables, in order to explain the local correlations between A and B. At first sight, it is very hard to conceive of such a cause in the past, given that measurement settings at remote locations are chosen at will by independent observers in the future. This is why it seems that such a solution requires an exception from the "Free Will" criterion. Either we need to invoke super-determinism from the past [45], or retro-causal effects of the future on the past [46,47], or "everything, everywhere, all at once" global determinism [16]. For a long time, it did not seem possible to find a classical explanation for quantum correlations, given their ability to violate Bell-type inequalities. Though, as shown above, a radically new solution is now available.

The *missing piece of this puzzle* is that mutually exclusive properties cannot exist at the same time in classical physics. Therefore, they cannot coincide objectively. Correlations between such events require subjective interventions. Someone needs to decide how many events to combine into a "pseudo-joint" measurement, leading to alternative scenarios with different coefficients of correlation. As seen in the "wheel-of-fortune" demonstration above, all the events are produced by a common physical mechanism – a spinning table. They are predetermined and recorded in fixed order. Measurement results cannot change after detection and obviously cannot influence each other in any way. Strictly speaking, we have the same $\lambda$ for all the events prior to detection. However, we need to qualify the mutual exclusivity of individual events with additional parameters $\lambda_x$, such as to define their manifestation at different points in time:

$$\lambda_A = \lambda + \lambda_1, \tag{8}$$
$$\lambda_B = \lambda + \lambda_2. \tag{9}$$

Thus, we have a novel explanation for the physical meaning of Equation (7). The two parameters $\lambda_A$ and $\lambda_B$ do not have to represent entirely different causal mechanisms. They can also correspond to a single shared causal process prior to observation, with sudden differentiation at the time of measurement. The nuance is that mutually exclusive properties cannot coincide, and yet another type of hidden variables is required. Essentially, we need to distinguish event-level (objective) hidden variables from correlation-level (subjective) hidden variables.

As shown above, we can choose to include all the four observables of a CHSH test ($A_1$, $B_1$, $A_2$ and $B_2$) in a single window of coincidence. In this case, we erase the temporal distinguishability between the four observables, and Equation (6) applies. This means that Bell violations are impossible. However, we can also choose to count the same observables in pairs, preserving the incompatible nature of recorded events. In this case, Equation (7) applies and maximal Bell violations become possible. Both of these patterns are available at the same time, with the same record of events. There is no "sub-ensemble post-selection" in this case, because *all* the events are considered in their order of detection. Therefore, we do not need super-classical mechanisms for explanation, because there is nothing physical to explain. The underlying reality does not need to change when observers decide to switch between different strategies for analysis. This aspect was overlooked in classical probability theory, and in the interpretation of Bell's Theorem.

From an intuitive point of view, it seems that objective Reality should be sufficient to explain any kind of correlation. A representative sample is a fixed record of observations. It is processed with a deterministic formula, such that the coefficient is uniquely determined. Hence, the traditional rule of thumb is:

$$1 \text{ Reality} \Rightarrow 1 \text{ coefficient.} \tag{10}$$

Nonetheless, mutually exclusive properties cannot coincide in the same context. For this reason, they cannot have unique coefficients of correlation. Instead, we have an observer-defined process that combines events from different contexts (across space and/or time). In this case, there is a forking path between reality and coefficients of correlation. A window of coincidence can be wider or narrower. It can start at one phase of a cycle or another. Other conditional factors are possible as well. Hence, the special rule of thumb for this case is:

$$1 \text{ Reality} \Rightarrow \text{multiple coefficients.} \tag{11}$$

Notably, the coefficients of correlation are not fully independent from objective considerations. Different contexts can be connected to each other by continuous transformations, or by adequate combinations in repeating cycles. There is a mechanism behind these correlations. However, physical reality does not automatically determine the combinations that might be selected for measurement. The actualization of one out of many possible correlations is an observer-dependent (and therefore subjective) process. We can summarize this conclusion with the slogan:

$$\textit{Objective relations, subjective correlations.} \tag{12}$$

Previously, it seemed impossible to reconcile quantum correlations with classical realism. Given the "1 coefficient" principle, it seemed that Reality itself is fundamentally subjective, when quantum coefficients are measurement dependent. In contrast, we see now that relationships between coefficients are not sufficiently explained by relationships between events. For a long time, quantum monogamy was used to define the gap between quantum and classical probability. Surprisingly, it has now become the key for their unification into a single deterministic framework.

## 5. Discussion

Quantum correlations do not have analogues in classical probability theory. The current attitude is that non-classical physical mechanisms are required for such patterns. Yet, there is a mismatch between classical probability theory and classical physics. Kolmogorov-type families of variables are restricted to jointly distributed configurations. This corresponds to predetermined properties that manifest *at the same time*. However, it is logically possible for predetermined properties to also occur at different times. For example, a set of mutually exclusive properties can manifest sequentially, in a pre-defined order. Similarly, some systems could experience mutually exclusive (but deterministic) transformations. In both cases, we can have predetermined measurement outcomes that cannot be analyzed with jointly distributed variables. Accordingly, it is still an open question whether quantum correlations can (or cannot) be explained with classical physics. Surprisingly, as shown above, the answer was hiding in plain sight. Mutually exclusive properties cannot coincide. The only way to conduct joint measurements is by combining events from different contexts or systems. This process of combination has unsuspected degrees of freedom, leading to multiple coefficients of correlation for the same recorded phenomena. Indeed, it is misleading to describe joint observations with the word "coincidence" in this case. Perhaps, they should be described as "pseudo-coincidences" or some other concept to be chosen in the future. It seems important to disambiguate the subjective combinations of incompatible events from the objective coincidences that are usually studied by classical probability theory.

The original goal of this study was to discover the "secret" behind quantum monogamy. How is it possible for a complete record of events to produce different coefficients of correlation at the same time? Why should it matter if two properties are counted at a time, or more? The answer, as shown above, is that incompatible properties do not have unique mappings between events and correlations. When different observables are paired across space or time, the simple choice between looking at two qualities or more can have automatic effects on the coefficients of correlation. Therefore, "quantum monogamy" does not reflect an objective change in the behavior of entangled quanta, but rather an artifact of "pseudo-correlation" between mutually exclusive properties.

Furthermore, this discovery leads to straightforward answers for several other open questions in quantum theory. In particular, an old theorem by Vorob'ev entails that Bell violations can be local in a classical sense [40,41]. In contrast, a more recent theorem by Shimony, Horne and Clauser [48] suggested the opposite: Bell violations are not locally possible without super-determinism. The solution is to realize that correlations for incompatible properties require supplementary hidden variables. (See *Appendix A* below). Therefore, they do not reflect relationships between objective event distributions. Both theorems are mathematically correct, and the apparent physical conflict is resolved. Local Bell violations do not contradict the possibility of "observer free will". Instead, they can emerge as a direct consequence of the free choice between various strategies for joint observation, even after detection, but only in the case of mutually exclusive physical properties.

A similar solution is available for another theoretical puzzle. Bell's definition of locality is based on the concept of Reichenbach separability. If two events do not influence each other directly, they should have a common cause. Therefore, their joint probability should be separable by virtue of a single "hidden variable". This is the basis for Bell's Theorem [19]. Yet, a subsequent theorem by Fine produced a strange result: Bell's inequality is only valid for jointly distributed variables [42]. By implication, it does not hold for mutually exclusive properties. How can this be? Why should it matter if properties exist at the same time or not, when they have a common cause in the past? As it turns out, mutually exclusive properties do not have fixed coefficients of correlation. Coincidences require additional hidden variables, and do not reflect objective relationships between events directly. Thus, it is possible for incompatible properties to violate "classical" separability, even if the underlying events are independent from each other. This is why Bell's theorem is mathematically correct, but the conjectures about local realism are not justified.

This conclusion leads to an important question. Monogamy of entanglement is an essential feature of quantum key distribution protocols. Emerging communication technologies (and especially their security) depend on this phenomenon. Is the future of quantum cryptography suddenly undermined? The simple answer is: "No". Quantum correlations are used to measure and quantify entanglement, but they never required explicit physical effects in space or time. The nature of Bell violations has always been a mystery. What we know now is that quantum correlations are still "subjective" in essence, without specific ontological requirements (such as action at a distance). Thus, it is not possible to falsify quantum correlations with traditionally defined "classical" means. Instead, the complexities of quantum correlations can now be absorbed by a new branch of classical probability theory, dedicated to mutually exclusive observations. More importantly, quantum states are still immune to cloning, they are still governed by uncertainty relationships, and they still express all the aspects of superposition. The most likely impact is that quantum communications will become more secure (not less), given that future designs can be informed by a deeper intuitive understanding of this phenomenon, with less reliance on trial and error.

The caveat is that non-commuting quantum variables are very different from the sequential properties that are analyzed above. More research is needed to explore this difference. Though, it is already clear that "super-quantum" correlations are possible in classical systems, and by extension in quantum systems too. Hence, it might be possible to identify cyclic quantum systems and to achieve correlations that go beyond the famous Tsirelson bound [49]. In this case, even the meaning of "distributed entanglement" would have to be reconsidered. It might be possible for system A to be maximally entangled with system B *and* alternatively with system C, so long as they are not all measured at the same time. Entanglement (defined as the ability to violate Bell-type inequalities) is not objectively monogamous, at least in principle. The known limits on shareability are subjective (*i.e.,* observation dependent), due to the relational nature of incompatible correlations.

In conclusion, it is true that classical systems *cannot* violate Bell's inequality. Though, it is also true that they *can* violate it with incompatible variables. This is not a contradiction, because a fixed record of events can have more than one coefficient of correlation. The same measurement outcomes, detected in the same order, produce coefficients that are "inseparable", as well as coefficients that are "separable". Therefore, statistical dependence does not necessarily correspond to physical dependence between events. By solving the mystery of quantum monogamy, we found a new way to think about quantum entanglement and a new way to explore its physical implications.

**Funding:** This research received no external funding.

**Data Availability Statement:** No database was created during this project.

**Acknowledgments:** This discovery was made possible by numerous debates over the years at the *Quantum Foundations Conference Series,* hosted by Linaeus University. The author is grateful to J.-A. Larsson, A. M. Cetto and A. Khrennikov for their help in understanding the nature of quantum statistics in ideal experiments. F. Čop provided valuable feedback on earlier drafts and editorial support. R. Jalba gave insightful feedback for *Discussion* section.

## Appendix: Correlation Without Causation for Hidden Variables

The goal of this study was to understand the nature of quantum monogamy. Yet, a side-effect of this solution is a new perspective on a different open problem in modern physics. It is known that Bell violations are possible with statistical independence between measurement outcomes and remote settings:

$$P(A|a,b) = P(A|a). \tag{A1}$$

This is known as the "no-signaling" condition. Furthermore, an old theorem by Vorob'ev [40,41] implied that such violations are possible for consistent families of variables, if they have cyclic arrangements. Indeed, CHSH experiments require cycles of overlapping pairwise measurements [50]. Therefore, it is theoretically possible for Bell violations to be non-signaling and local at the same time. Nonetheless, quantum correlations are currently described as examples of "non-signaling non-locality" [37,38]. That is because Bell violations entail contradictory coefficients of correlation. If we assume a 1-to-1 correspondence between event distributions and their correlations, then contradictory coefficients presuppose contradictory physical conditions for each joint observation. This was clarified with a theorem by Shimony, Horne and Clauser (SHC) in 1976 [48]. Thus, Alice's events may not measurably depend on the settings of Bob, if they are non-signaling. Yet, the hidden variables that determine Alice's events must be correlated with Bob's remote settings, given the inconsistency between different coefficients. This is why "local" explanations of Bell violations are currently presumed to require super-determinism [45]. Nonetheless, we have just seen that the mentioned "1-to-1 correspondence" between events and correlations does not exist for mutually exclusive properties. Instead, joint measurements require additional hidden variables that correspond to the fluid process of creating recorded coincidences for objectively non-coincident events. Hence, the conflict between the theorems of SHC and Vorob'ev is resolved by noting the independence of event-level hidden variables from correlation-level hidden variables. Bell violations require contradictory coefficients of correlation, but this is not necessarily reducible to contradictions between outcomes, or between the conditions for their manifestation.

Let us consider a simple example. Suppose that Alice and Bob make an experiment in which Alice can choose between observing $A_1$ *or* $A_2$, while Bob is measuring $B_1$ *and* $B_2$ in the same iteration. Furthermore, suppose that Bob's events are always correlated ("+,+" or "−,−") if Alice chooses $A_2$, but always anticorrelated ("+,−" and "−,+") if Alice chooses $A_1$. It seems obvious that Alice's decision works like a switch. She can control at will if Bob gets correlated or anti-correlated coincidences for the same observables. In any case, we have a strong correlation between the measurement settings of Alice and the presumed hidden variables that explain the pattern of coincidence for Bob. Nonetheless, if we look at the "wheel-of-fortune" toy model from Figure 1, we see the same pattern at work without any magic. Any Alice event ($A_1$ or $A_2$) is flanked by two different Bob events ($B_1$ and $B_2$). Every Alice event can be used in two different pairings with Bob, dictating which neighbors fall into the same window of coincidence. Yet, no influences are possible between events, because their values are written down in advance. Instead, $A_1$ events are always flanked by opposite values for Bob's observables, while $A_2$ events are not.

On closer inspection, Alice's choice of measurement does not force the events of Bob to change correlations. It is the other way around: Alice's events can only be found in one region of the "wheel-of-fortune" or another, with predetermined patterns of potential coincidence. Just as shown above for quantum monogamy, this is a combinatorial effect. Any time we have a cluster of positive values, followed by a cluster of negative values, some events are going to be in the "middle of the pack" with identical neighbors, while others are going to be "at the edge". In this case, $A_1$ events are always between anticorrelated values for $B_1$ and $B_2$, while $A_2$ events have correlated neighbors. Consequently, the apparent "action at a distance" between Alice and Bob is non-physical.

Remarkably, there is no post-selection in this case. All the possible measurement outcomes of Bob are counted exhaustively. Yet, they happen at regular intervals, like clockwork. The same cycle of 4 events can be broken down at will into correlated or anti-correlated pairs, simply by displacing the starting point of the window of coincidence (Figure A1). This process can be guided by using $A_1$ or $A_2$ values as anchors. In other words, correlations between hidden variables and remote measurement settings do not correspond to physical influences at the event level. All the observables emerge from the same causal mechanism (a spinning table) and therefore require a single configuration of hidden variables prior to detection. Instead, alternative strategies of joint measurement can lead to different coefficients of correlation for the same record of events.

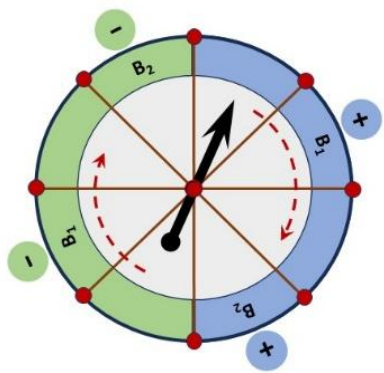

**Figure A1.** The outcomes for $B_1$ and $B_2$ are spaced at equal intervals on the "wheel of fortune". The 4 possible events can be separated arbitrarily into correlated pairs (on the left and right halves of the table) or anticorrelated pairs (on the top and bottom halves). Both patterns can be isolated from the same record of events, by anchoring the window of coincidence on either $A_1$ or $A_2$ measurements.

To sum up, "joint measurements" do not reveal an objective feature of reality in this case. Instead, they correspond to an observer-defined method for creating artificial pairings between objectively non-coincident events. Such behavior may seem "non-classical", but only if the starting assumptions are deficient. Even when there is a clear formal dependence between hidden variables at one place and measurement settings at another, correlation is not proof of causation.